\documentclass[12pt,preprint]{aastex}

\slugcomment{}

\shorttitle{Haro~3}
\shortauthors{}

\begin{document}

\title{Revealing the Young Starburst in Haro~3 with Radio and Infrared Imaging}


\author{Kelsey E. Johnson \altaffilmark{1}, R\'emy Indebetouw, Christer Watson}
\affil{Department of Astronomy, University of Wisconsin,
    Madison, WI 53706} 
\and

\author{Henry A. Kobulnicky}
\affil{Department of Physics and Astronomy, University of Wyoming, 
Laramie, WY 82071}


\altaffiltext{1}{NSF Astronomy \& Astrophysics Postdoctoral Fellow}

\begin{abstract}

\end{abstract}
The Wolf-Rayet galaxy Haro 3 (Mrk~35, NGC~3353) was observed at the
near-IR and radio wavelengths as part of ongoing program to study the
earliest stages of starbursts.  These observations confirm that the
current episode of star formation is dominated by a single region
(region~A).  While there are knots of recent ($\sim 10$~Myr) star
formation outside of region~A, the sources of ionizing radiation as
observed in both radio and Br$\gamma$ observations are almost
exclusively associated with region~A.  The derived ionizing flux
implies a star formation rate of $\approx 0.6 M_\odot$~yr$^{-1}$
localized within a radius of $\sim 0.1$~kpc.  A comparison with
observations from HST indicates that one or more of the star clusters
in region~A are optically obscured.  The star clusters in region~A
have ages at least as young as $\sim 5$~Myr, and possibly as young as
$\sim 0.1$~Myr.  The star cluster that appears to be the youngest
also exhibits a near-IR excess in its colors, possibly indicating
natal dust in very close proximity to the ionizing stars.  The
difference between optical- and radio-determined ionizing fluxes as
well as the near-IR colors indicate an average extinction value of
$A_V \approx 2.5$ in region~A.  The total stellar mass associated with
the current starburst in region~A is inferred from both the near-IR
and radio observations to be $\sim 10^6 M_\odot$.  The other main
stellar concentrations observed in the near-IR (Regions~B1 and B2) are
somewhat older than region~A, with ages $\sim 8-10$~Myr, and the
near-IR observations indicate they have stellar masses of $\sim 8
\times 10^4$ and $\sim 2 \times 10^4 M_\odot$, respectively.


\section{INTRODUCTION}
The study of nearby starburst galaxies is important for advancing our
understanding of the evolution of stars and galaxies.  Understanding
the nature of starburst systems becomes increasingly important at
higher redshifts where merger events and the resulting starburst
episodes are common.  According to hierarchical models of structure 
formation, galaxy mergers are responsible for the various 
galaxy morphologies we observe in the local universe, and these mergers
may also contribute to the reionization of the universe at $z>5$
\citep[e.g.][]{madau99}. 

The early stages of a starburst episode are typically obscured by the
molecular clouds associated with the star formation phenomenon.
Although newly born massive stars emit optical and ultraviolet (UV)
light prodigiously, observations at these wavelengths short-ward of the
infrared (IR) are of limited use in studying star {\it formation}.  In
order to mitigate the effects of extinction and obtain accurate
measurements of the current star formation in a galaxy, observations
in the infrared to radio regimes are critical.

The infrared emission from starburst regions is primarily due to two
sources: (1) blackbody emission from the stars themselves, and (2)
thermal re-radiation by dust of the optical and UV emission from the
stars.  Which of these sources dominates depends largely on the
wavelength being observed; the mid- to far-IR spectral energy
distribution is dominated by dust emission, while the near-IR emission
is typically due to stellar light.  Dust emission can also contribute
to the near-IR spectral energy distribution if it resides in very
close proximity to the stars and reaches temperatures near those
required for sublimation.  In this case, the near-IR emission is
observed to have an ``infrared excess''.  The near-IR colors
of clusters can be used in combination with population
synthesis models to determine properties of the embedded stellar
content.

While broad-band near-IR observations can directly measure the stellar
photospheric emission, radio observations probe the embedded \ion H2
regions surrounding massive young stars.  Young massive stars are
responsible for producing thermal free-free emission via their
surrounding \ion H2 regions.  If an \ion H2 region is young and dense
enough, this thermal emission will actually be optically thick in the
cm regime and exhibit a turnover in the spectral energy distribution.
The radio flux densities from \ion H2 regions can be used to infer the
total stellar content, age, electron densities, and pressures of
extremely young starburst regions.

Haro~3 is a dwarf irregular starburst galaxy that was selected for
infrared and radio study based on its relatively close proximity of
13.1~Mpc and known Wolf-Rayet (WR) stellar content \citep{steel96}.
Haro~3 has an optical diameter of 3.8~kpc and consists of two main
sites of star formation, A and B \citep{steel96}, as indicated in
Figure~\ref{XcontV}a.  Its classification as a WR-galaxy indicates
that the system has undergone a starburst episode within the last
$\sim 3-6$~Myr, making it an excellent candidate for hosting sites of
ongoing star formation less than a few Myr old.  \citet{steel96} find
a WR/O-star ratio of 0.31, suggesting the recent starburst episode was
relatively ``instantaneous'' and estimate a burst age of $\sim
1.5-3$~Myr for region~A and $\sim 5$~Myr for region~B.  Infrared Space
Observatory (ISO) observations by \citet{metcalfe96} indicate that
polycyclic aromatic hydrocarbon (PAH) emission peaks on the main
regions of star formation, but is present throughout the galaxy.  This
global PAH emission indicates copious near-ultraviolet emission
throughout the galaxy, and provides additional evidence for recent
wide-scale star formation in Haro~3.

The cause of the current vigorous burst of star formation in Haro~3 is
unclear.  There is circumstantial evidence of a minor merger event,
including both extended emission that is suggestive of tidal tails and
two distinct regions of star formation that could be remnants of the
original galaxies.  However, \citet{steel96} note that the galaxy has
relatively regular outer isophotes in optical light and that all of
the star forming regions have the same redshift.  \citet{steel96} also
find approximately the same metallicity for both regions A and B of
$12+log(O/H)=8.4$ (roughly 1/3 the solar value); using a different
slit location, \citet{huang99} find a metallicity of
$12+log(O/H)=8.3$.  This evidence leads Steel et al. to conclude that
the star formation in Haro~3 is most likely self-induced.  We discuss
various star formation scenarios for Haro~3 in \S~\ref{speculation}.

\section{OBSERVATIONS}
\subsection{Radio Imaging}
Haro~3 was observed with the Very Large Array (VLA) 
\footnote{The National Radio Astronomy Observatory is a facility of the 
National Science Foundation operated under cooperative agreement by 
Associated Universities, Inc.}
at 3.6~cm using
the A-array in February 2002 and at 1.3~cm using C-array in November
2002.  Both observations utilized the flux density calibrator 3C286;
based on the scatter in the VLA flux calibrator database, we estimate
the resulting flux density scale at each wavelength is uncertain by
$\lesssim 10$\%.  Calibration was carried out using the AIPS software
package, and the data sets were inverted and cleaned using the task
{\sc imagr}.  While the total uv-coverage observed at each wavelength
is different, an attempt was made to achieve relatively well-matched
synthesized beams by restricting the uv-range used in the imaging
process to 50-250~k$\lambda$ at each wavelength.  Although the
uv-sampling within this range varied between the wavelengths, the
resulting images are sensitive to roughly the same spatial scales.
The weighting schemes used in the imaging process were also varied in
order to mitigate the effect of different uv-sampling within the
restricted uv-range.  An additional 3.6~cm image was made using the
full uv-range of $\sim 20 - 1000$~k$\lambda$ in order to achieve the
maximum resolution and sensitivity possible.  The resulting parameters
for these images are listed in Table~\ref{imaging}.  Flux densities
were measured using identical apertures within the {\sc viewer}
program in AIPS++; several combinations of apertures and annuli were
used in order to estimate the uncertainty in this method.  The resulting
flux densities and their uncertainties are listed in Table~\ref{fluxes}.

\subsection{Near-Infrared Imaging}
Near-infrared (near-IR) imaging of Haro~3 was obtained in January 2003
using the NIRIM camera \citep{meixner99} on the WIYN 3.5m telescope at
Kitt Peak National Observatory.  Images were taken using the
broad-band J (1.257$\mu$m), H (1.649$\mu$m), and K' ($2.12\mu$m)
filters as well as narrow-band Br$\gamma$ (2.166$\mu$m) with a plate
scale of $0.19''$/pixel and a field-of-view of $\sim 0.8'$.  A
chopping technique was used to remove atmospheric background; short
exposures ($< 30$s) were taken in a off-off-on-on-off-off pattern.
While the natural seeing as measured from standard stars was excellent
throughout the observations ($\sim 0.3 - 0.5''$), the combination of
the short exposures obtained in the chopping process resulted in a
degradation of the imaging quality due to the telescope pointing error
and lack of point sources to use for registration in the Haro~3 field.
The resulting image quality is $\sim 0.7''$.  The total on source
integration times for J, H, K', and Br$\gamma$ were 480, 240, 240, and
720~s, respectively.  A set of infrared standard stars were observed
at different airmasses throughout the night roughly every hour.  The
data were reduced and calibrated using the IDL and IRAF software
packages.  The median of the four off-source images surrounding each
on-source image was subtracted from the on-source image, and the
resulting on-source images were cross-correlated and combined.  Based
on the uncertainties in the photometric solutions, we estimate the
flux calibration to be accurate within $\sim 10$\%.  This accuracy was
verified by a different project carried out on the same night that
contained stars present in the 2MASS catalog \citep{indebetouw03}.
Aperture photometry was performed using the IRAF package {\sc DAOPHOT}
with identical apertures and annuli in each of the three broad-band
filters.

\section{RESULTS}

\subsection{The Starburst Morphology in the Optical, Near-IR, and Radio}
\begin{figure}
\plottwo{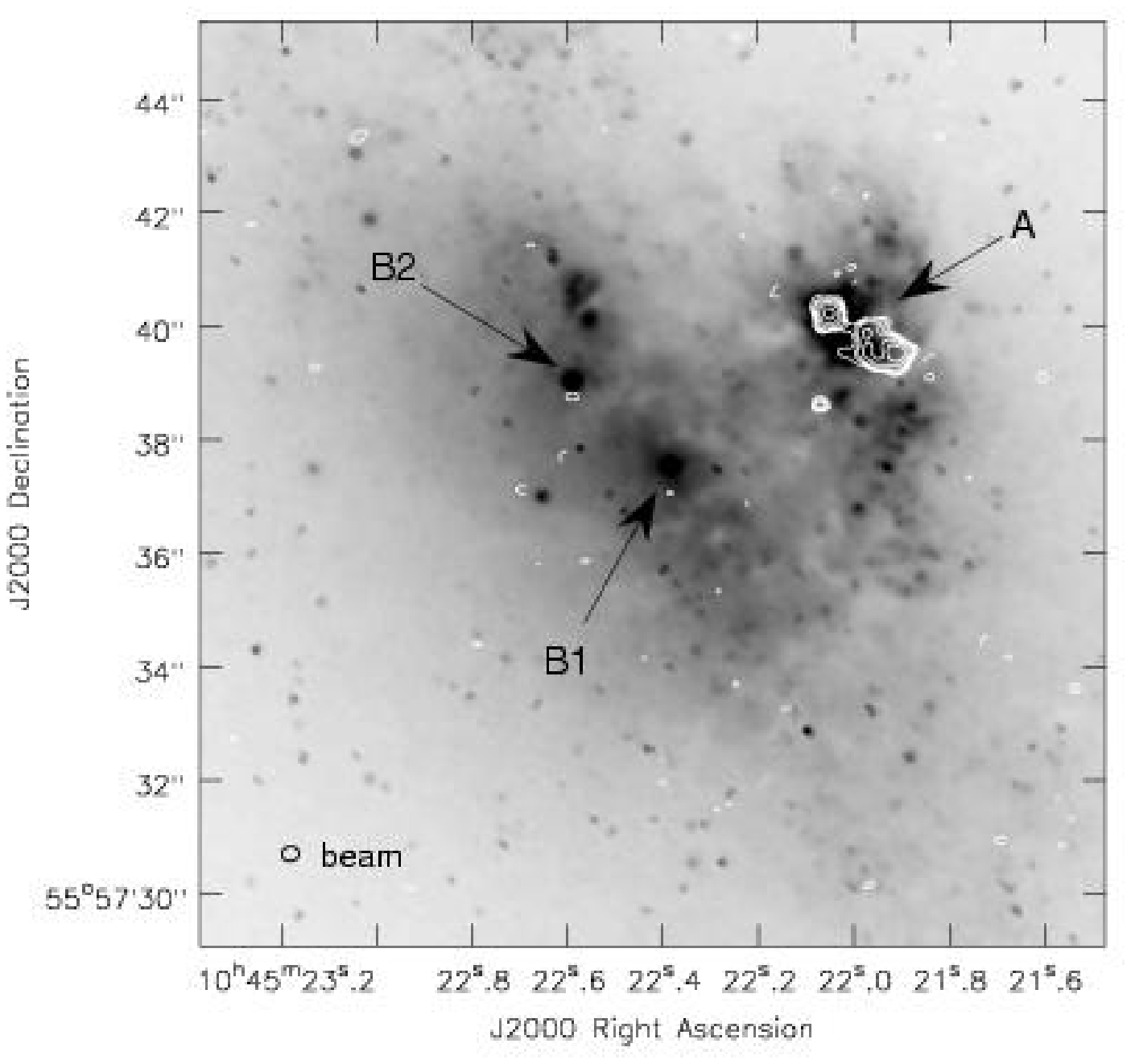}{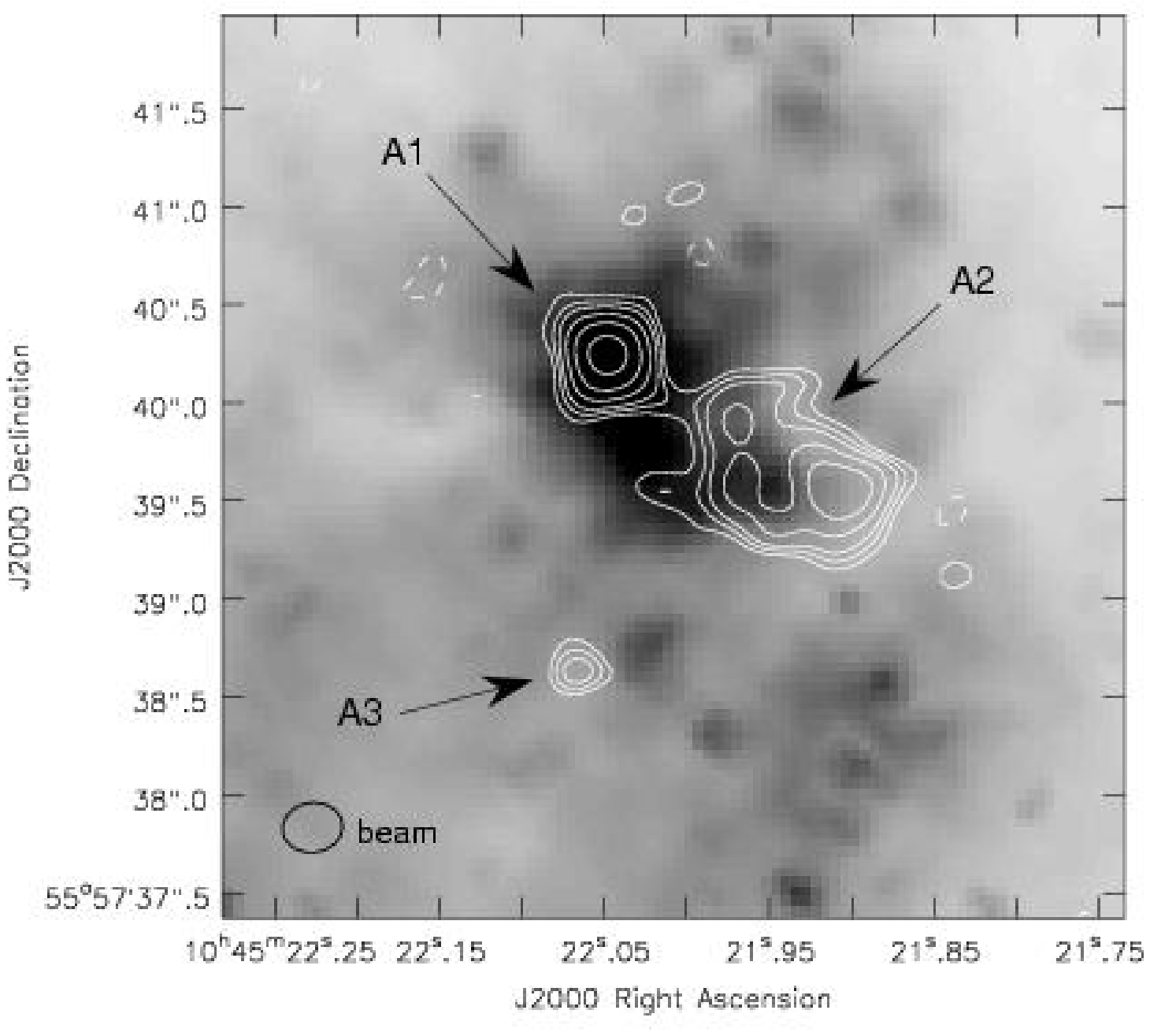}
\caption{a) VLA 3.6~cm contours of Haro~3 overlaid on and HST V-band
gray scale image. The contour levels are $-3,3,4,5,7,10,15 \times \sigma$
(0.023~mJy/beam). b) An enlargement of region~A.  \label{XcontV}}
\end{figure}

\begin{figure}
\plottwo{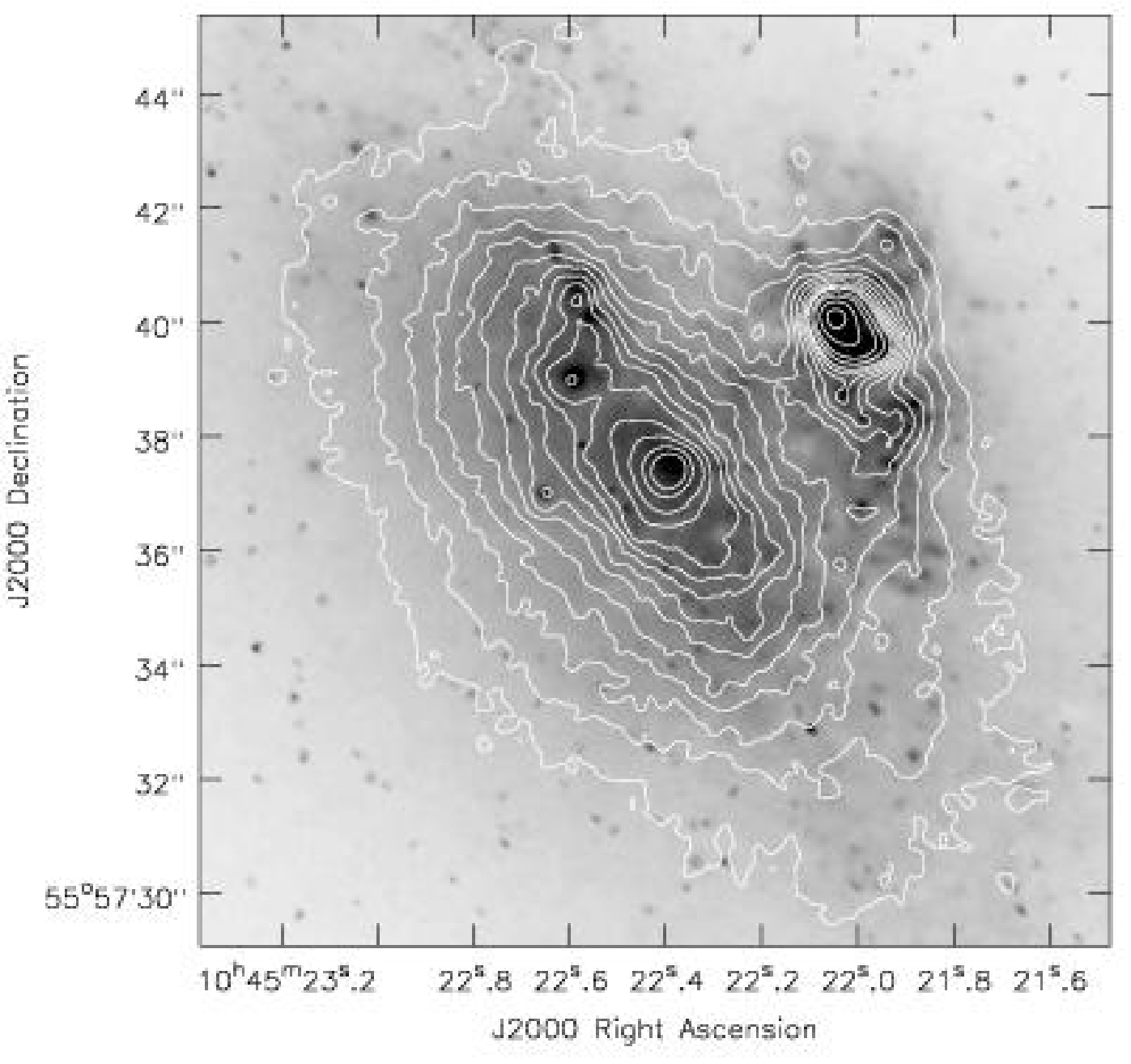}{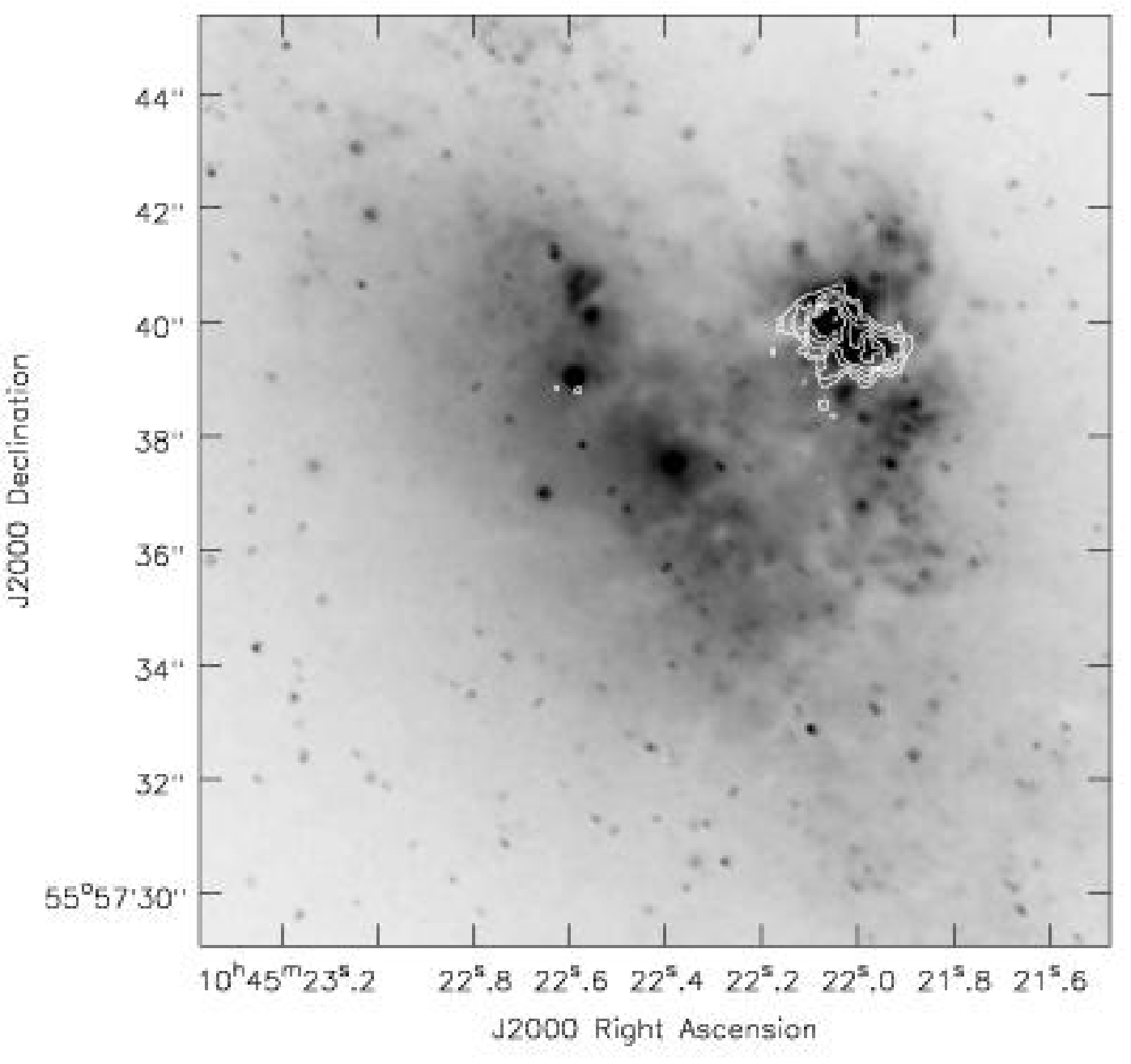}
\caption{NIRIM Near-IR contours overlaid on the HST V-band 
gray scale image. a) J-band contours, b) Br$\gamma$ contours.  \label{JcontV}}
\end{figure}

A V-band (F606W) optical image was retrieved from the Hubble Space
Telescope (HST) archive for comparison with the radio and near-IR
observations presented in this paper.  Unfortunately, only a single
``snapshot'' image was available, and therefore the PSF is
under-sampled and the image suffers from cosmic ray contamination.
This image is shown along with radio and near-IR contours in
Figures~\ref{XcontV}-\ref{JcontV}.  The Regions A and B clearly
contain a number of optically bright starburst knots, many of which
are likely to be super star clusters (SSCs).  The optical image also
suggests a number of dust lanes between regions A and B and to the
west of region A.

The radio emission at 1.3~cm and 3.6~cm is thermal in nature 
($S_\nu \propto \nu^\alpha$, where $\alpha \gtrsim 0$) and {\it
only} associated with region~A, which supports the idea that region~A
is dominating the current burst of star formation in the galaxy
system.  This thermal radio emission is resolved into at least three
components at both 1.3~cm and 3.6~cm, which we call A1, A2, and A3.
The highest resolution 3.6~cm image also suggests that A2 is complex
in nature, consisting of a single dominant source and two or more less
luminous sources (Fig.~\ref{XcontV}b).  The relative registration of
the optical and radio images are limited by the astrometric
uncertainty of HST, which we estimate to be $\lesssim 1''$.  The radio
source A1 corresponds to the brightest optical starburst knot, however
source A2 extends beyond the optical starburst region into a possible
dust lane and does not appear to be associated with an optical source.
Likewise, source A3 is located in the proximity of possible dust
lanes, but could possibly be associated with an optical source if the
images were shifted relative to each other by $\sim 0.3''$.  However,
within the registration uncertainty, only one of the three radio
sources (A1, A2, or A3) could have an optical counterpart.

The near-IR images were registered to the optical image using two
stars common to both images which are outside the field-of-view shown
in Figure~\ref{JcontV}.  We estimate this registration between the
optical and the near-IR to have an accuracy better than $\sim 0.2''$.
As shown in Figure~\ref{JcontV}, most of the optically bright
starburst knots are associated with peaks in the near-IR emission.  In
all of the near-IR bands, the peak emission is associated with source
A1.  Although there is an extension of this near-IR emission in the
direction of the radio source A2, there is not a peak in near-IR
emission associated with it.  There appears to be near-IR emission
associated with the radio source A3, however the peak in the emission
appears to shift slightly ($\lesssim 0.3''$) depending on the near-IR
band: while a peak in the K-band emission is in excellent spatial
agreement with the radio peak, the J-band peak is more closely
associated to the optical source offset by $\sim 0.3''$ to the west of
the radio source.  This relative morphology may indicate that source
A3 is complex and has not undergone coeval star formation throughout
throughout the object or suffers from variable forground extinction.
The optical regions B1 and B2 are also clearly associated with peaks
in the near-IR emission in all of the broad bands.  The
continuum-subtracted nebular Br$\gamma$ emission is almost exclusively
associated with region~A, although there is a hint of Br$\gamma$
emission from region~B2.  The Br$\gamma$ observations confirm the
youth and strength of the starburst in region~A as inferred from the
radio observations.

\subsection{Ionizing Luminosities and Star Formation Rate \label{ionizing}}

The ionizing luminosity of starburst regions is an important quantity in
their physical interpretation; from the ionizing flux, both the
star formation rate and the mass of the stellar content can be
estimated.  In the case of these observations, both the thermal radio 
luminosity and the Br$\gamma$ emission can be used to independently 
determine the ionizing flux of the starburst regions in Haro~3.  

The production rate of Lyman continuum photons can be determined from 
the radio luminosities by following \citet{condon92},
\begin{equation}
{{Q_{Lyc}}}
\geq6.3\times10^{52}~{{\rm s}^{-1}}\Big({{T_e}\over{10^4~{\rm K}}}\Big)^{-0.45}
\Big({{\nu}\over{{\rm Ghz}}}\Big)^{0.1}
\Big({{L_{thermal}}\over{10^{27}~{\rm erg~s^{-1}~Hz}^{-1}}}\Big).
\end{equation}
We adopt a ``typical'' \ion H2 electron temperature of $T_e
= 10^4$K, and use the 1.3~cm luminosities (which are less likely to
suffer from either self-absorption or non-thermal contamination than
the 3.6~cm luminosities).  The $Q_{Lyc}$ values for sources A1, A2,
and A2 determined using this method are listed in Table~\ref{fluxes},
and they range from $Q_{Lyc} = 0.3 - 2.1 \times 10^{52}$~s$^{-1}$.
The entire region~A has an ionizing luminosity inferred from the radio
observations of $Q_{Lyc} = 5.7 \times 10^{52}$~s$^{-1}$

The continuum-subtracted Br$\gamma$ fluxes can be used to determine
$Q_{Lyc}$ by following \citet{ho90},
\begin{equation}
{{Q_{Lyc}}}
= 2.9\times 10^{51}~{{\rm s}^{-1}} \Big({{D}\over{\rm Mpc}}\Big)^{2}
\Big({{3 F_{Br\gamma}}\over{{10^{-12}{\rm erg~s}^{-1}{\rm cm}^{2}}}}\Big).
\end{equation}
Unfortunately, it is not possible to disentangle sources A1 and A2 in
the Br$\gamma$ observations.  We present the $Q_{Lyc}$ values for
regions A1+A2, A3, and B2 using this method in
Table~\ref{Br_gamma}. The inferred ionizing flux for region~A using
its Br$\gamma$ flux is $Q_{Lyc} = 3 \times 10^{52}$~s$^{-1}$, roughly
$2\times$ lower than the $Q_{Lyc}$ value determined above using the
radio observations.  This difference in ionizing flux as measured from
radio and near-IR observations suggests that even the Br$\gamma$
emission is suffering from some extinction by the natal material
surrounding the young clusters; the factor of two suggests $A_K
\approx 0.8$ or $A_V \approx 8$ for the entire region~A.  This value
is higher than the extinction value inferred from near-IR colors for
the individual regions A1-A3 in \S~\ref{extinctions}, possibly indicating 
that region~A contains areas with thermal radio emission that are completely
obscured in the near-IR.

The ionizing luminosity can also be converted to a star formation rate (SFR).
Following \citet{kennicutt98}, 
\begin{equation}
{{SFR(M_\odot {\rm year}^{-1})}}
= 1.08\times 10^{-53}~Q_{Lyc}({{\rm s}^{-1}}).
\end{equation}
Using the ionizing flux measured from the radio observations, region~A
has a star formation rate of $\sim 0.6 M_\odot$~yr$^{-1}$ within a radius
of less than 0.1~kpc.  This is roughly twice the value of 
$\sim 0.34 M_\odot$~yr$^{-1}$ estimated by \citet{hunter82} from the H$\beta$ 
emission for the {\it entire} galaxy.  This difference in derived star 
formation rates underscores the need for long wavelength observations in 
order to detect the emission from the current regions of star formation.  
This is particularly true for dwarf galaxy systems in which a single 
region of embedded star formation can have a dramatic impact on the 
inferred star formation rate.
%an extremely high star formation density of
%$\sim 5 M_\odot$~yr$^{-1}$~kpc$^{-2}$.
%is higher than the star formation rates for any of the dwarf irregular
%galaxies in the \citet{hunter85} sample.  
With a total atomic and molecular gas mass of $\approx 6.7 \times 10^8
M_\odot$ \citep{meier01,gordon81}, the gas depletion time scale for
Haro~3 at the current star formation rate is roughly 1~Gyr, which is
atypically short for a dwarf irregular galaxy.  This high star
formation rate suggests that Haro~3 is undergoing an atypical star
formation event for a dwarf galaxy, and might lend support to the idea
that this galaxy is undergoing a small-scale merger event.

\subsection{Star Cluster Extinctions, Ages, and Masses \label{extinctions}}
\begin{figure}
\centerline{
\rotatebox{-90}{
\resizebox{6in}{!}{{\includegraphics{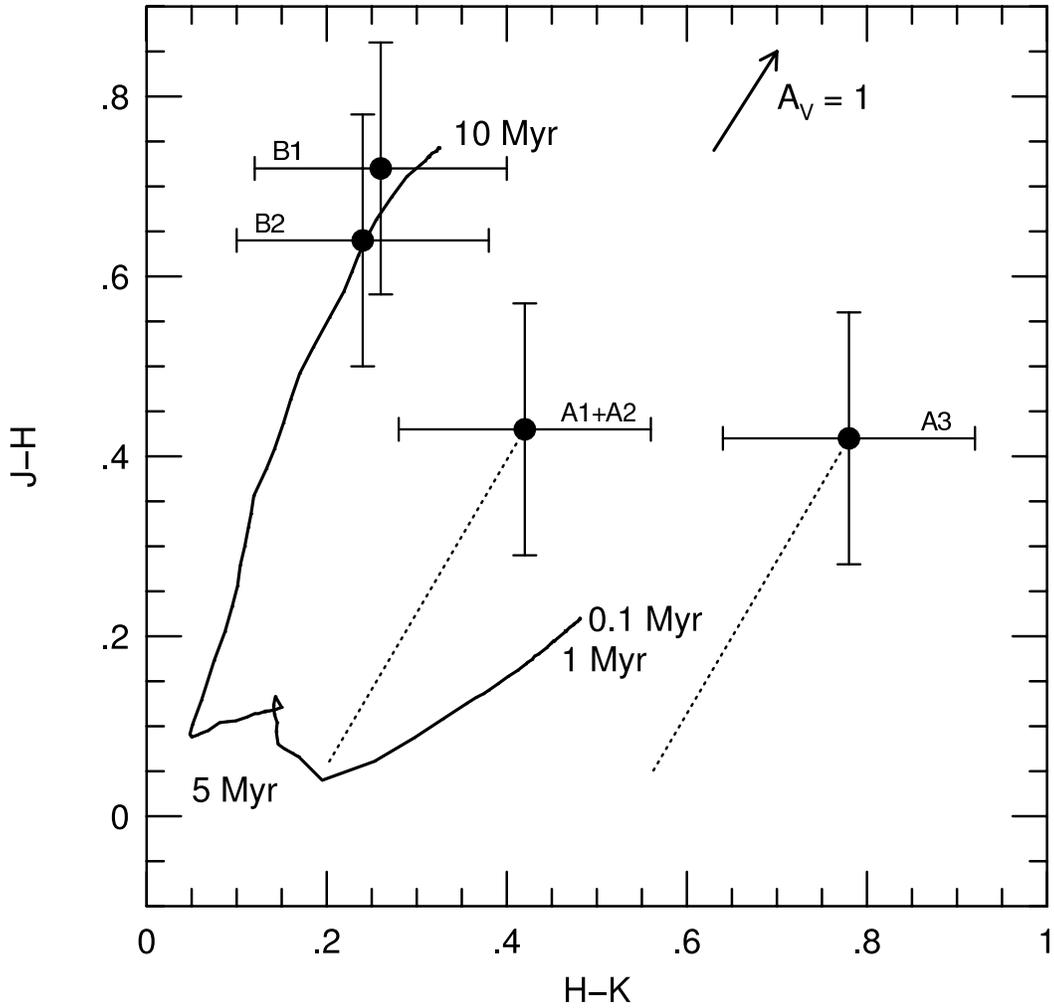}}}}}
\caption{Near-IR color-color plot of sources in Haro~3.  A model track 
from 0.1~Myr to 10~Myr is overlaid \citep{leitherer99}.  Sources A1 and A2
could not be resolved from each other in the near-IR images and are shown 
as a single point in this plot. In order to facilitate comparison with 
the model, a de-reddening projection is indicated for sources A1+A2 and A3.
 \label{J_HvsH_K}}
\end{figure}

In order to estimate the physical properties (such as extinction, age,
and mass) of the near-IR sources in Haro~3, we have adopted the
Starburst99 population synthesis models of \citet{leitherer99}.  In
this case, we use models with $Z=1/2 Z_\odot$ and a Salpeter initial
mass function between $1-100 M_\odot$.  It is well known that the
stochastic presence of asymptotic giant branch (AGB) stars can affect
the integrated light of a stellar population due to both their
luminosity and red colors \citep[e.g.][]{charlot96}; however, because
objects A1, A2, and A3 are young enough to have associated thermal
radio emission, it is likely that they are far too young to also host
AGB stars.  A color-color plot of the near-IR sources is shown along
with the model track for cluster ages of 0.1-10~Myr in
Figure~\ref{J_HvsH_K}.  Sources A1 and A2 were not resolved from each
other in the near-IR images and are referred to collectively as A1+A2.
In order to be consistent with the model for any age, sources A1+A2
and A3 must have extinction values of $A_V \sim 2-4$.  These values
are lower than those implied for the entire region~A from a comparison
between the ionizing flux as measured with radio and Br$\gamma$
emission in \S~\ref{ionizing} (and discussed in \S~\ref{discussion}).

When these sources are de-reddened and projected onto the model
track (shown with the dotted lines in Figure~\ref{J_HvsH_K}), source
A1+A2 must be younger than $\sim 5$~Myr, and source A3 has an H-K
color that is redder than even the youngest age available in these
models of 0.1~Myr.  The offset of A3 from the model is consistent with
an infrared excess \citep{lada92}, possibly indicating very hot dust
in the vicinity of the stellar cluster.  Sources B1 and B2 are in
excellent agreement with the model track for ages of $\sim 8-10$~Myr
and no extinction.  However, while these sources are consistent with
being unreddened, we cannot rule out extinction values as high as $A_V
\sim 5$.  In any case, sources B1 and B2 must be older than $\sim
5$~Myr to be in agreement with the model.

The mass of the stellar clusters can be estimated using either their
near-IR luminosities or their ionizing fluxes (derived in
\S~\ref{ionizing}) along with the Starburst99 models (assuming an
instantaneous burst with a Salpeter IMF, 100~$M_\odot$ upper cutoff,
1~$M_\odot$ lower cutoff, and metallicity of $1/2 Z_\odot$).  In the
first method, the K-band magnitudes can be used in combination with
the estimated age of a cluster to determine its mass.  Adopting an age
of $\sim 0-2$~Myr for sources A1+A2 and A3, these objects have stellar
masses of $M_{*} \approx 1 \times 10^6 M_\odot$ and $M_{*} \approx 2
\times 10^4 M_\odot$, respectively.  Adopting an age of $\sim
8-10$~Myr for sources B1 and B2 yields stellar mass of $M_{*} \approx
8 \times 10^4 M_\odot$ and $M_{*} \approx 2 \times 10^4 M_\odot$,
respectively.  The second method can be utilized for the radio sources
given their inferred ionizing fluxes in combination with the
Starburst99 models.  Assuming that $Q_{Lyc}$ scales directly with the
cluster mass and that the clusters are $\lesssim 2$~Myr old, we infer
that the radio sources A1, A2, and A3 have stellar masses of $\sim
3.4\times 10^5 M_\odot$, $5.2\times 10^5 M_\odot$, and $4\times 10^4
M_\odot$ respectively.  These masses are in excellent agreement with
the values derived using the K-band luminosity.  The entire starburst
region~A has an inferred stellar mass from the radio observations of
$9.2\times 10^5 M_\odot$; this estimate does not include {\it recent}
star formation ($t \gtrsim$ a few Myr) that is no longer detected
at radio wavelengths.

\section{DISCUSSION \label{discussion}}

\subsection{Comparison of Optical, Near-IR, and Radio Results}
The results from observations of Haro~3 across several decades in
wavelength are forming a consistent picture of the star formation in
the dwarf system over the last several million years.  Previous work
by \citet{steel96} indicated a number of star forming knots, with
region~A being the youngest with an estimated age of $\sim 2.5$~Myr.
This age is consistent with the upper limit we derive from near-IR
observations of $< 5$~Myr for regions~A1 and A2 and $< 1$~Myr for
region~A3.  The near-IR and radio observations presented in this paper
also yield similar stellar masses of $\sim 10^6 M_\odot$ for region~A.

A comparison between the extinction values measured by observations at 
different wavelengths may provide some insight into the
geometry of the star forming regions.  The extinction value we infer
for region~B $(A_V \sim 0)$ is consistent with the extinction value
calculated by \citet{steel96} using Balmer line ratios; however, the
extinction values for region~A are discrepant.  Steel et al.\
calculate an $A_V \sim 0.2$, while the near-IR colors presented here
indicated $A_V \sim 2-4$.  Moreover, using the H$\beta$ flux, Steel et
al. calculate that region~A has an ionizing flux of $Q_{Lyc} \approx 5
\times 10^{51}$~s$^{-1}$.  This value is roughly 10 times lower than
the value inferred from our radio observations, indicating an
extinction value of $A_V \approx 2.5$.  To complicate the extinction 
estimate further, a comparison between the ionizing flux from region~A 
measured from radio and Br$\gamma$ observations in this work suggests
an $A_V \sim 8$.

The different extinction values inferred from diagnostics at different
wavelengths might be expected if the dust geometry is not that of a
simple screen between us and the obscurred cluster.  Three possible
scenarios are the following: (1) If the gas and dust are mixed,
observations at a given wavelength only probe the material to an
optical depth of $\tau \approx 1$, and consequently yield different
extinction estimates.  While Balmer line emission can only be detected
from regions with little reddening, Br$\gamma$ observations probe to
even greater depths, and radio emission does not suffer from
extinction at all.  (2) If the the dust is clumpy \citep[as might be
expected in a medium of the type advocated by][]{faison98}, a fraction
of optical flux can leak out of regions that otherwise suffer from
large amounts of obscuration.  (3) The region used to measure the
extinction with Balmer lines by Steel et al.\ contained all of
region~A, including a great deal of recent star formation that is
optically visible (see Fig.~\ref{XcontV}b).  This ``size of aperture''
effect will also tend to produce an apparently lower extinction value.
The radio and near-IR observations preferentially detect the youngest
sources that are still likely to be obscured by larger amounts of
dust.

Steel et al.\ note that the low extinction value they measure for
Haro~3 is ``interesting'' considering that this galaxy is a bright
IRAS source, suggesting a large warm dust content.  Haro~3 also merits
comparison with the dwarf starburst galaxy Henize~2-10 in this regard;
Haro~3 has $log(L_{FIR})=9.57$ \citep{soifer87}, and Henize~2-10 has
$log(L_{FIR}/L_\odot)=9.73$ \citep{vacca02}, both of which are
extremely high values for dwarf galaxies.  In the case of Henize~2-10,
only four embedded SSCs are responsible for most of the mid-IR flux of
this galaxy, despite the fact that it also contains roughly 80
optically visible SSCs \citep{vacca02, johnson00}.  Likewise, the warm
dust surrounding the current regions of star formation is likely to
be localized to small regions within Haro~3 and not uniforming
distributed throughout the galaxy.

\subsection{Importance of the IR Excess}

Young stellar objects in the Milky Way are often observed to have an
IR-excess 
%in their H-K color \citep[e.g.][]{lada92}.  
in their JHK colors \citep[K flux brighter than the locus of reddened
main sequence stars in a color-color diagram,][]{lada92}.  This
IR-excess is commonly interpreted as being due to extremely hot dust
(2000-3000K) present at the inner edges of circumstellar disks.  In
the case of young massive star clusters, an IR-excess may also suggest
the presence of hot circumstellar material.  Similarly red H-K colors
have also been observed by \citet{devost99} for clusters in Arp~299
and by Buckalew, Dufour, \& Kobulnicky (in prep) in a sample of WR
galaxies.

In principle, Br$\gamma$ flux can also contribute to a red H-K color;
however, in the case of cluster A3 in Haro~3, the Br$\gamma$ flux is
not high enough to appreciably contribute to the K-band magnitude.  A
uniform underlying older stellar population is also not likely to
contribute to the IR-excess because the background subtraction removes
any bias toward an underlying color.  Another possibility to consider
is that a single evolved star (e.g. AGB or carbon star) is
contaminating the observations.  While these types of stars can have
both K-band magnitudes and H-K colors consistent with those of source
A3, they would have J-H colors that are much redder (J-H$>1$) than any
of the objects presented in this paper \citep{wainscoat92}.
Furthermore, we would not expect such a star to also have radio
continuum and Br$\gamma$ emission.  While we are not claiming with
certainty that the IR-excess observed in these clusters is due to
circumstellar material, it appears to be the most plausible
explanation.  Moreover, we know that young star clusters must contain
hot circumstellar dust early in their evolution, so we should not be
surprised to find it.  We predict that nearly all radio selected
ultra-young clusters will have an IR-excess in follow-up near-IR
observations.

\subsection{Speculation on the Cause of the Starburst in Haro~3 
\label{speculation}}

The origin of starburst episodes in dwarf galaxies has been a subject
of much discussion in the literature.  These galaxies tend to be
dynamically ``simple''; they often lack spiral structure and show
evidence of solid body rotation \citep[e.g.][]{gallagher84}.  These
dynamic features suggest that star formation cannot be due to
triggering processes such as gas compression by density waves or
shear.  \citet{gerola80} put forward a model of stochastic
self-propagating star formation (SSPSF) that naturally explains
starburst episodes in isolated dwarf galaxies as a statistical
fluctuation in the star formation rate.

While SSPSF is almost certainly contributing to the star formation in
in Haro~3 on some level, it seems unlikely that it could be solely
responsible for the current burst for two main reasons: (1) The
starbursts in regions~A and B cannot be causally related.  Regions~A
and B are approximately 250pc apart, and a shock traveling at a speed
of $\sim 10$~km~s$^{-1}$ would take approximately 25~Myr to traverse
this distance.  However, the bursts in regions~A and B are too similar in age
to allow for this scenario; region~A is roughly 1-4~Myr old,
and the burst in region~B is roughly 8-10~Myr old (and stellar
feedback would not have been an important factor for at least a few
Myr until the first supernovae began to explode in region~B).  (2) In
order to form a single SSC, energetics require the {\it convergence}
of at least $\sim 10^{51}$~ergs \citep{elmegreen04}.  Only previous
star formation in the form of SSCs could generate this amount of
energy, and SSPSF would result in the energy {\it diverging}.
A single supernova shell or expanding bubble from an OB association
\citep[as proposed by][]{steel96} would be far from adequate.

In order to trigger the formation of SSCs, \citet{elmegreen04} argues
that star formation must be triggered by one of three physical
environments: (1) instabilities and turbulence on a kpc-scale; (2)
galaxy interactions or mergers; (3) galactic nuclear regions.  In the
case of Haro~3, we can almost certainly rule out (3) as the origin for
all of the SSCs as there is no obvious ``nuclear'' region.  It is
possible that morphology of the recent star formation in Haro~3
represents an underlying spiral structure, in which case (1) becomes
a possibility.  The evidence against (2) as presented by
\citet{steel96} is that regions~A and B have similar redshifts and the
outer {\it optical} isophotes of the galaxy are relaxed.  However, the
fact that regions~A and B have similar redshifts does not rule out the
possibility that they are (or once were) separate systems.  Moreover,
the relaxed outer {\it optical} isophotes of the system do not rule
out a small scale interaction such as that found in the dwarf
starburst galaxy Henize~2-10 \citep{corbin93, kobulnicky95, johnson00}.  

A number of comparisons can be drawn between Haro~3 and Henize~2-10 in
regard to their possible interaction history.  Like Haro~3,
Henize~2-10 appears to be isolated and has relaxed outer optical
isophotes; however, Henize~2-10 appears to be ``interacting'' with a
massive cloud of gas that is falling into the main body of the galaxy
\citep{kobulnicky95}.  The HI spectral profile of He~2-10 provided by
Kobulnicky et al. is smooth and single peaked, with a possible {\it
slight} asymmetry.  However, in the HI map of He~2-10 with a
resolution of $\approx 30''$, there is a clear protrusion of the HI
that is consistent with a tidal tail.  In the HI observations of
Haro~3 \citep{gordon81}, the HI spectral profile is similar in nature
to that of He~2-10; it has a smooth, single-peaked structure and
possibly a {\it slight} asymmetry.  However, these observations had a
resolution of only $\approx 10'$; in the absence of a high resolution
map, this spectral profile might support the status of Haro~3 as
non-interacting.  However, even at this low resolution, Gordon \&
Gottesman note that the HI in Haro~3 has ``faint irregular
nebulosity''.  Higher spatial resolution maps of the HI in Haro~3
might enable us to conclusively determine whether a small-scale
interaction is the origin of the current starburst in galaxy.  Moreover,
high resolution HI observations similar to those used to study the dwarf
galaxy Holmberg~II by \citet{puche92} would allow for a detailed study of 
shells and bubbles associated with the recent starburst episodes in Haro~3.

\acknowledgments

We thank Jay Gallagher, Bill Vacca, Peter Conti, and Brent Buckalew
for useful discussions related to this paper.  The anonymous referee
provided a number of comments that led to improvements in the
manuscript.  We also thank Pat Knezek and Alan Watson for valuable
help with the NIRIM camera.  K.E.J. gratefully acknowledges support
for this paper provided by NSF through an Astronomy and Astrophysics
Postdoctoral Fellowship.  Support for proposal \#09934 was provided by
NASA through a grant from the Space Telescope Science Institute, which
is operated by the Association of Universities for Research in
Astronomy, Inc., under NASA contract NAS5-26555.

\clearpage

\clearpage

\clearpage

\clearpage
\begin{deluxetable}{lccccc}
\tabletypesize{\scriptsize}
\tablecaption{Imaging Parameters \label{imaging}}
\tablewidth{0pt}
\tablehead{
\colhead{$\lambda$}&\colhead{Weighting}&\colhead{UV-Range}&
\colhead{Synth. Beam}&\colhead{P.A.}&\colhead{RMS noise}\\
\colhead{(cm)}&\colhead{(robust value)}&\colhead{(k$\lambda$)}&
\colhead{($''\times ''$)}&\colhead{($\degr$)}&\colhead{(mJy/beam)} }

\startdata
1.3 & 0 & 50-250 & $0.88\times 0.78$ & -86 &0.05\\
3.6 & 5 & 50-250 & $0.70\times 0.62$ & -88 & 0.05\\
3.6 & 5 & 20-1000 & $0.31\times 0.25$ & -83 & 0.02\\
\enddata
\end{deluxetable}

\begin{deluxetable}{lcccccc}
\tabletypesize{\scriptsize}
\tablecaption{Properties of Haro~3 Radio Sources \label{fluxes}}
\tablewidth{0pt}
\tablehead{
& \colhead{$F_{1.3cm}$} & \colhead{$F_{3.6cm}$} & \colhead{$Q_{Lyc}$} & 
\colhead{M$_{stars}$} & \\

\colhead{Source} & \colhead{(mJy)} & \colhead{(mJy)} &
\colhead{($\times 10^{52}$s$^{-1}$)} & \colhead{($\times 10^5 M_\odot$)} &
\colhead{$\alpha^{3.6cm}_{1.3cm}$} }

\startdata
A1 & $1.18\pm0.12$ & $1.05\pm0.15$ & 2.1 & 3.5 & $0.12\pm0.18$\\
A2 & $1.78\pm0.18$ & $1.55\pm0.16$ & 3.2 & 5.3 & $0.14\pm0.15$\\
A3 & $0.15\pm0.02$ & $0.14\pm0.02$ & 0.3 & 0.5 & $0.07\pm0.20$\\
\enddata
\end{deluxetable}

\begin{deluxetable}{lcccccc}
\tabletypesize{\scriptsize}
\tablecaption{Properties of Haro~3 Near-IR Sources \label{Br_gamma}}
\tablewidth{0pt}
\tablehead{
& & \colhead{M$_{stars}$} & \colhead{$F_{Br\gamma}$} & \colhead{$Q_{Lyc}$}&  \\

\colhead{Region} & \colhead{$M_K$} & \colhead{($\times 10^5 M_\odot$)} 
& \colhead{(mJy)} & \colhead{($\times 10^{52}$s$^{-1}$)} 
}

\startdata
A1+A2  & $-16.1\pm0.1$ & 11  & $1.4\pm0.4$ & 2.8  \\
A3     & $-11.8\pm0.2$ & 0.2 & $0.1\pm0.05$ & 0.2 \\
B1     & $-14.7\pm0.1$ & 0.8 & ---          & --- \\
B2     & $-12.9\pm0.2$ & 0.2 & $0.1\pm0.05$ & 0.2 \\
\enddata
\end{deluxetable}


\begin{thebibliography}{}

\bibitem[Charlot, Worthey, \& Bressan(1996)]{charlot96} 
Charlot, S., Worthey, G., \& Bressan, A. 1996, \apj, 457, 625

\bibitem[Condon(1992)]{condon92} Condon, J.J. 1992, \araa, 1992, 30, 575

\bibitem[Corbin, Korista, \& Vacca(1993)]{corbin93}
Corbin, M.R.., Korista, K.T., \& Vacca, W.D. 1993, \aj, 105, 1313

\bibitem[Devost(1999)]{devost99} Devost, D. 1999, \aj, 118, 549

\bibitem[Elmegreen(2004)]{elmegreen04} Elmegreen, B. 2004, in 
preparation for the proceedings of ``The Formation and Evolution of 
Massive Young Star Clusters'', ASP, eds. L. Smith and H. Lamers 

\bibitem[Faison et al.(1998)]{faison98}
Faison, M., Churchwell, E., Hofner, P., Hackwell, J., Lynch, D.K., \& 
Russell, R.W. 1998, \apj, 500, 280

\bibitem[Gallagher \& Hunter(1984)]{gallagher84} 
Gallagher, J.S. \& Hunter, D.A. 1984, \araa, 22, 37

\bibitem[Gerola, Seiden, \& Schulman(1980)]{gerola80} 
Gerola, H., Seiden, P.E., \& Schulman, L.S. 1980, \apj, 242, 517

\bibitem[Gordon \& Gottesman(1981)]{gordon81} Gordon, D. \& Gottesman, S. 
1981, \aj, 86, 161

\bibitem[Ho, Beck, \& Turner(1990)]{ho90}
Ho, P.T.P., Beck, S.C., \& Turner, J.L. 1990, \apj, 349, 57

\bibitem[Huang et al.(1999)]{huang99}
Huang, J.H., Gu, Q.S., Ji, L., Li, W.D., Wei, J.Y., \& Zheng, W. 1999, 
\apj, 513, 215

\bibitem[Hunter, Gallagher, \& Rautenkranz(1982)]{hunter82}
Hunter, D.A., Gallagher, J.S., \& Rautenkranz, D. 1982, \apjs, 49, 53

\bibitem[Indebetouw et al.(2003)]{indebetouw03} Indebetouw, R.,
Watson, C., Johnson, K.E., Whitney, B., \& Churchwell, E. 2003, \apjl,
596, 83

\bibitem[Johnson et al.(2000)]{johnson00} Johnson, K.E., Leitherer, C., 
Vacca, W.D., \& Conti, P.S. 2000, \aj, 120, 1273

\bibitem[Kennicutt(1998)]{kennicutt98} Kennicutt, R.C. 1998, \araa, 36, 189

\bibitem[Kobulnicky et al.(1995)]{kobulnicky95} Kobulnicky, H.A., Dickey, J.M.,
  Sargent, A.I., Hogg, D.E., \& Conti, P.S. 1995, \aj, 110, 116 

\bibitem[Lada \& Adams(1992)]{lada92} Lada, C.J. \& Adams, F.C. 1992, 
\apj, 393, 278

\bibitem[Leitherer et al.(1999)]{leitherer99} Leitherer, C., Schaerer,
D., Goldader, J.D., Delgado, R.M.G., Robert, C., Kune, D.F., de Mello,
D.F., Devost, D., Heckman, T.M. 1999, \apjs, 123, 3

\bibitem[Madau, Haardt, \& Rees(1999)]{madau99}
Madau, P., Haardt, F., \& Rees, M.J. 1999, \apj, 514, 648

\bibitem[Meier et al.(2001)]{meier01}
Meier, D.S., Turner, J.L., Crosthwaite, L.P., Beck, S.C. 2001, \aj, 121, 740

\bibitem[Meixner, Young Owl, \& Leach(1999)]{meixner99}
Meixner, M., Young Owl, R., \& Leach, R.W. 1999, \pasp, 111, 997

\bibitem[Metcalfe et al.(1996)]{metcalfe96} Metcalfe, L., Steel, S.J., 
Barr, P., Clavel, J., Delaney, M., Gallais, P., Laureijs, R.J., Leech, K., 
McBreen, B., Ott, S., Smith, N., \& Hanlon, L. 1996, \aap, 315, 105

\bibitem[Puche et al.(1992)]{puche92} Puche, D., Westfall, D., Brinks, E., 
\& Roy, J. 1992, \aj, 103, 1841

\bibitem[Soifer et al.(1987)]{soifer87}
Soifer, B.T., Sanders, D.B., Madore, B.F., Neugebauer, G., Danielson, G.E., 
Elias, J.H., Lonsdale, C.J., \& Rice, W.L. 1987, \apj, 320, 238

\bibitem[Steel et al.(1996)]{steel96} Steel, S.J., Smith, N.,
Metcalfe, L., Rabbette, M., McBreen, B. 1996, \aap, 311, 721

\bibitem[Vacca, Johnson, \& Conti(2002)]{vacca02}
Vacca, W.D., Johnson, K.E., \& Conti, P.S. 2002, \aj, 123, 772

\bibitem[Wainscoat et al.(1992)]{wainscoat92}
Wainscoat, R.J., Cohen, M., Volk, K., Walker, H.J., \& Schwartz, D.E. 1992, 
\apjs, 83, 111 


\end{thebibliography}
\end{document}